%Paper: hep-th/9204049
%From: BERSHADSKY@huhepl.harvard.edu
%Date: Wed, 15 Apr 1992 23:15:05 -0400 (EDT)

\input harvmac

\def\IB{\relax{\rm I\kern-.18em B}}
\def\IC{\relax\hbox{$\inbar\kern-.3em{\rm C}$}}
\def\IP{\relax{\rm I\kern-.18em P}}
\def\IR{\relax{\rm I\kern-.18em R}}
\def\IN{\relax{\rm I\kern-.18em N}}
\def\IZ{\relax\ifmmode\mathchoice
{\hbox{Z\kern-.4em Z}}{\hbox{Z\kern-.4em Z}}
{\lower.9pt\hbox{Z\kern-.4em Z}}
{\lower1.2pt\hbox{Z\kern-.4em Z}}\else{Z\kern-.4em Z}\fi}

\def\brs{{\scriptscriptstyle\rm BRST}}

\def \CT{{\cal T}}

\def\CM{{\cal M}}

\def \CV{{\cal V}}

\def\frac#1#2{{#1\over#2}}
\def\coeff#1#2{{\textstyle{#1\over #2}}}
\def\hf{\coeff12}

\def\journal#1&#2(#3){\unskip, \sl #1\ \bf #2 \rm(19#3) }
\def\andjournal#1&#2(#3){\sl #1~\bf #2 \rm (19#3) }

\Title{PUPT--1315, HUTP--92/A016}
{{\vbox {\centerline{Scattering of Open and Closed Strings}
\medskip
\centerline{in 1+1 Dimensions}
\bigskip
}}}

\centerline{M. Bershadsky}
\medskip\centerline
{\it Lyman Laboratory,}
\centerline{\it Harvard University,}
\centerline{\it Cambridge, MA 02138}
\bigskip
\centerline{and}
\bigskip
\centerline{D. Kutasov}
\medskip\centerline
{\it Joseph Henry Laboratories,}
\centerline{\it Princeton University,}
\centerline{\it Princeton, NJ 08544.}

\vskip .2in

\noindent
The ground ring structure of 1+1 dimensional string theory leads to
an infinite set of non linear
recursion relations
among the `bulk' scattering amplitudes of open and closed
tachyons on the disk,
which fix them uniquely.
The relations are generated by the action of the ring on the tachyon
modules;
associativity of this action determines all structure constants.
This algebraic structure may allow one to relate the
continuum picture to a matrix model.

\Date{4/92}
%\draftmode

\newsec{Introduction.}

An outstanding problem in two dimensional string theory
(see e.g.
\ref\kaz{V. Kazakov, Paris preprint LPTENS-90/30 (1990).},
\ref\kleb{I. Klebanov, Princeton preprint PUPT-1271 (1991).},
\ref\KUS{
D. Kutasov, Princeton preprint PUPT-1277 (1991).}\ for reviews)
is to reconstruct the matrix model, which is a very powerful and
natural
description of the physics of the theory directly from the continuum
approach. It has been suggested
\ref\wkms{E. Witten,
IAS preprint HEP-91/51 (1991).}\ (see also
\ref\klpol{I. Klebanov and A. Polyakov, Mod. Phys. Lett. {\bf A6} (1991)
3273.},
\ref\wz{E.Witten and B. Zwiebach, IAS preprint IASSNS-HEP-92/4 (1992).})
that certain spin zero, ghost number zero states which exist
in this model and generate the so called `ground ring', are important
in this context. In this interpretation, the ground ring is
identified with the ring of functions on the phase space of the
matrix model. The physical excitations
of the theory, the massless `tachyons', correspond to infinitesimal
perturbations of the Fermi surface, a certain curve on phase space.
$W_\infty$ type symmetries of phase space are generated by certain
dimension
$(1,0)$, $(0,1)$ currents. In fact, the whole structure, some aspects
of which we will review below, is superficially very similar to the
matrix
model picture developed in
\ref\dj{S. Das and A. Jevicki, Mod. Phys. Lett. {\bf A5} (1990) 1639;
A. Sengupta and S. Wadia, Int. Jour. Mod. Phys. {\bf A6} (1991) 1961;
D. Gross and I. Klebanov, Nucl. Phys. {\bf B352} (1991) 671.},
\ref\polc{J. Polchinski, Nucl. Phys. {\bf B362} (1991) 125.},
\ref\greg{G. Moore, Nucl. Phys. {\bf B368} (1992) 557; G. Moore,
R. Plesser and S. Ramgoolam, Yale preprint P35-91 (1991).},
but
so far the program of deriving the detailed structure of \dj, \polc, \greg\
from the
continuum has not been completed.

While it may seem that the issue is of academic interest only since
an
exact solution exists, the real interest in this subject lies
in understanding other vacua of two dimensional string theory for
which
the continuum picture is much more developed (either because
a matrix model description is lacking or because the matrix model
has so far proven too hard to solve). Examples include the black hole
\ref\black{E. Witten, Phys. Rev. {\bf D44} (1991) 314.},
two dimensional fermionic string theory \KUS,
\ref\DIIFK{P. Di Francesco and D. Kutasov, Princeton preprint
PUPT-1276 (1991), Nucl. Phys. {\bf B}, in press.}\ and superstring theory
\ref\sup{D. Kutasov and N. Seiberg, Phys. Lett. {\bf 251B} (1990)
67.},
and open plus closed $2d$ string theory
\ref\bk{M. Bershadsky  and D. Kutasov.
Phys. Lett. {\bf 274B} (1992) 331.} (see also
\ref\tan{Y. Tanii and S. Yamaguchi, preprint STUPP-92-128 (1992).}).
In these situations, the analogs of the ground ring exist, and it
would
be nice to be able to use them to work up to an exact solution and/or
to an underlying matrix model.

In this paper we are going to investigate open plus closed $2d$ string
theory
on the disk along the lines of the above comments. In a previous
paper
\bk\ we have found the bulk \KUS\ amplitudes for scattering of open
string `tachyons' (again massless) on the disk by explicitly
evaluating
certain Veneziano integrals. Here, we will start by showing that
the result follows quite elegantly from the ring relations on the
tachyon modules \wz,
\ref\KMS{D.Kutasov, E. Martinec and N. Seiberg, Phys. Lett. {\bf 276B}
(1992) 437}.
We will then go on to consider the case of arbitrary disk scattering
amplitudes with any number of closed and open string tachyons, which
are difficult to evaluate directly, and find that they satisfy
an infinite number of non-linear relations which again follow from
the action of the ring on the modules.
Compatibility of these relations determines all structure constants
involved and all correlation functions.
In fact the situation is infinitely over constrained, so it's
not guaranteed apriori that any solutions exist (nevertheless, this
is
of course the case).

The action of the ring on the modules is again reminiscent
of the phase space structure of a matrix model. We conclude the
paper
by some observations regarding this relation, postponing further
discussion to future work. But first, in the next section we start,
to set the stage, by briefly describing some general properties
of bulk amplitudes in $D=d+1$ dimensional
open $+$ closed string theory, and review the two dimensional
(closed) case on the
sphere.

\newsec{Some preliminaries.}

\subsec{Veneziano amplitudes in $d+1$ dimensions.}

Open string states first come into play on world sheets with the
topology of a disk (which we will denote by $\CM$). The appropriate
action of  world sheet $2d$ gravity is (see e.g.
\ref\open{C. Callan, C. Lovelace, C. Nappi and S. Yost, Nucl. Phys.
{\bf B288} (1987) 525; R. Metsaev and A. Tseytlin, Nucl. Phys. {\bf
B298}
(1988) 109.}):
\eqn\action{\eqalign{
{\cal S}=&{1\over2\pi}\int_{\cal
M}d^2\xi\sqrt{g}\left[g^{ab}G_{\mu\nu}
(X)\partial_a X^\mu\partial_b X^\nu-R^{(2)}\Phi(X)+\CT(X)\right]\cr
+&{1\over\pi}\int_{\partial\cal M}d\xi\left[A_\mu(X)\partial_\xi
X^\mu
-K\Phi(X)+g^{1\over4}\CV(X)\right]\cr}
}
where $g_{ab}$ is the world sheet metric, $R^{(2)}$ its scalar
curvature, $g={\rm det} g_{ab}$,
and $K$ the extrinsic curvature on the boundary $\partial\cal M$.
$X^\mu$, $\mu=0,1,\cdots, d$ parametrize the space-time manifold, and
$G_{\mu\nu}$, $A_\mu$, $\Phi$, $\CT$, $\CV$ correspond to general
metric, gauge field, dilaton and closed and open
string tachyon condensates, respectively.

We will be mainly discussing here the ``linear dilaton'' vacuum of
\action:
\eqn\vac{G_{\mu\nu}=\eta_{\mu\nu};\;A_\mu=0;\;\Phi(X)=QX^0;\;
Q^2={25-d\over3}}
As is by now standard (see e.g.
\ref\sei{N. Seiberg, Prog. Theor. Phys. Suppl. {\bf 102} (1990) 319.},
\KUS, \DIIFK)
in a vacuum such as \vac, where the string coupling
$g_{\rm st}\simeq e^\Phi$ depends on $X^\mu$, we have to supply some
sort of
``wall'' to keep the system away from strong coupling, e.g. by
turning
on appropriate potentials
$\CT(X)$, $\CV(X)$. However, physics in the `bulk' of space-time
(not to be confused with the bulk of the world sheet) is insensitive
to such boundary effects, and can be studied by putting
$\CT(X)=\CV(X)=0$. This reduces the problem to free field theory
on the world sheet with anomalous conservation of $X^0$.

The emission vertices for closed and open string states are:
\eqn\clop{
\eqalign{
\CT_k(z,\bar z)=&\exp(ik\cdot X+\beta_k X^0);\;\; \hf
k^2-\hf\beta(\beta+Q)=1\cr
\CV_k(\sigma)=&\exp(ik\cdot X+\beta_k X^0);\;\; \hf
k^2-\hf\beta(\beta+
\hf Q)=1/4\cr}}
where we have conformally transformed the disk to the upper half
plane
$z=\sigma+i\tau$, $\tau\geq0$, with the open vertices inserted at
$\tau=0$.
The propagator on the upper half plane
(with Neumann boundary conditions) is:
$\langle X^\mu(z)X^\nu(w)\rangle=-\eta^{\mu\nu}\left(\log|z-w|^2
+\log|z-\bar w|^2\right)$.
Scattering of the tachyons \clop\ is described by the Veneziano
integral
representation (valid for $m\geq3$):
\eqn\venez{
\eqalign{
&\langle\CT_{k_1}\cdots\CT_{k_n}\CV_{p_1}\cdots\CV_{p_m}\rangle=
\prod_{i=1}^n\int d^2z_i\prod_{j=4}^m\int dw_j
\prod_{i=1}^n|z_i|^{4{\bf k_i\cdot p_1}}|1-z_i|^{4{\bf k_i\cdot
p_2}}\cr
&\prod_{1=i<j}^n|z_i-z_j|^{2{\bf k_i\cdot k_j}}
|z_i-\bar z_j|^{2{\bf k_i\cdot k_j}}
\prod_{1=i}^n|z_i-\bar z_i|^{{\bf k_i\cdot k_i}}
\prod_{j=4}^m|w_j|^{4{\bf p_j\cdot p_1}}|1-w_j|^{4{\bf p_j\cdot
p_2}}\cr
&\prod_{4=i<j}^n|w_i-w_j|^{4{\bf p_i\cdot p_j}}
\prod_{i=1,j=4}^{n,m}|z_i-w_j|^{4{\bf k_i\cdot p_j}}
\cr}}
where we have used the notation ${\bf k_1\cdot k_2}\equiv k_1\cdot
k_2-\beta_1
\beta_2$, and (anomalous) energy momentum conservation:
$\sum_{i=1}^nk_i+\sum_{j=1}^mp_j=0$,
$\sum_{i=1}^n\beta_i+\sum_{j=1}^m\beta_j=-Q/2$.
As usual, one has to fix the ordering of the boundary operators (up
to
cyclic permutations). Different orderings correspond to different
channels. The amplitude \venez\ is $SL(2,R)$ invariant; therefore we
have
the freedom to fix three open vertices on the boundary (as we did in
\venez\ ),
or one open vertex and one closed vertex operator.

The amplitude \venez\ exhibits poles corresponding to scattering of
closed
$\rightarrow$ closed, closed $\rightarrow$ open and open
$\rightarrow$
open strings. It is easy to check that they factorize on physical on
shell
states in all possible channels.
For general dimension $d$ one can not say much more about \venez\ due
to the
complicated dynamical structure of $d+1$ dimensional string theory.
In $1+1$ dimensions one can actually evaluate these amplitudes
and therefore in the next sections we will concentrate
on the situation in two space-time dimensions.

\subsec{Closed $2d$ strings on the sphere.}

Before passing to the case of interest to us here, the disk, we will
briefly review the structure on the sphere, in order to have
at hand a collection of relevant facts, to illustrate the logic and
to
compare later to the structure on the disk.

The system is described by \action, \vac\ with $\CM=$ sphere and
$Q=2\sqrt2$ (see \vac). We will consider here the case where $X^0, X^1$
are both non compact. The compact case can be treated along similar
lines.
The closed `tachyon' emission vertex  (in normalizations
of $\CT_k$ and $k$ which will be convenient later) is:
\eqn\closed{\CT_k^{(\pm)}=-{\Gamma(\pm{k\over2})\over
\Gamma(1\mp{k\over2})}
\exp\left[i{k\over2\sqrt2}X+(-\sqrt{2}\pm{k\over2\sqrt2})\phi\right]}
where $\CT_k$ has to be either multiplied by $c\bar c$
(reparametrization
ghosts) or integrated over $\CM$ for BRST invariance.
$\CT_k^{(+)}$ $(\CT_k^{(-)})$ describe right (left) moving massless
particles.
In addition to the tachyons \closed\ the spectrum includes an
infinite
set of `discrete' oscillator states at momenta $k\in 2\IZ$, which are
related
to the symmetries mentioned in the introduction; we will not need
those
below.
The ground ring is generated by the operators
\eqn\aclosed{a_{\pm}=-|cb -{1 \over \sqrt 2}(\partial \phi \mp
i\partial X)|^2
                        e^{\pm iX/\sqrt2 + \phi/\sqrt2 }}
which are BRST closed (but not exact) and have $\Delta=\bar\Delta=0$.
The ring is spanned by the set of BRST invariant operators
$\{(a_+)^n(a_-)^m\}$, $n,m\in\IZ_+$.

As pointed out in \KMS,
tachyon dynamics is constrained by the ring relations on the modules
\closed.
By using free field OPE, which is valid in bulk correlators \DIIFK,
we find (for $k\not\in2 \IZ$):
\eqn\mod{
\eqalign{
a_\pm \CT_k^{(\pm)}=&\CT_{k\pm2}^{(\pm)}\cr
a_\mp \CT_k^{(\pm)}=&0\cr}}
Both relations in \mod\ are true modulo BRST commutators, but while
the first survives in (bulk) correlation functions, the second
receives non-linear modifications.

Indeed, consider the amplitude:
\eqn\FZ{F(z,\bar z)=\langle a_-(z)c\bar c\CT_{k_1}^{(+)}(0)c\bar c
\CT_{k_2}^{(+)}(\infty)\prod_{i=3}^n\int \CT_{k_i}^{(+)}(z_i)
c\bar c\CT_p^{(-)}(1)\rangle}
(most other amplitudes can be shown to vanish \KUS,
\DIIFK). One can convince oneself
that $\partial_z F=\partial_{\bar z}F=0$ by using
$\partial_za_-=\{Q_\brs,b_{-1}a_-\}$ and deforming the contour
(commuting
$Q_\brs$ to the other operators in \FZ). This is not completely trivial --
one has to check vanishing of boundary terms from the $z_i$
integrals. Given that $F(z,\bar z)=F$ is constant one can derive
a non-linear identity on correlation functions of $\CT_k$ by
comparing $F(z\rightarrow1)$ and $F(z\rightarrow0)$. In the former
limit we get by the first equation in \mod:
$\langle\prod_{i=1}^n \CT_{k_i}^{(+)}\CT_{p-2}^{(-)}\rangle$
while in the second one would find $0$ by using \mod\ naively. In
fact,
there is a correction which can be schematically summarized\foot{Here
and in many subsequent formulae we ignore factors of $\pi$ which can
be
easily restored.}
as follows
\wz, \KMS,
\ref\igor{I. Klebanov, Princeton preprint PUPT-1302 (1992); S.
Kachru, Princeton
preprint PUPT-1305 (1992).}:
\eqn\nonlin{a_-\CT_{k_1}^{(+)}\int\CT_{k_2}^{(+)}=
\CT_{k_1+k_2-2}^{(+)}}
I.e., in the presence of an integrated $\CT_{k_2}$ the contribution
of the region
in which $a_-$ approaches $\CT_{k_1}$ is modified. It is easily checked
that \nonlin\ is the only modification of \mod\ which is necessary.
By combining \mod, \nonlin\ in \FZ\ we find a recursion relation:
\eqn\sprec{
\langle\prod_{i=1}^n\CT_{k_i}^{(+)}\CT_{p-2}^{(-)}\rangle
=\sum_{i=3}^n\langle\CT_{k_1+k_i-2}^{(+)}\prod_{i\not=j=2}^n\CT_{k_j}
^{(+)}\CT_p^{(-)}\rangle}
whose solution (given $\langle
\CT_{k_1}^{(+)}\CT_{k_2}^{(+)}\CT_p^{(-)}
\rangle=1$) is:
\eqn\cl{
\langle\prod_{i=1}^n\CT_{k_i}^{(+)}\CT_p^{(-)}\rangle=(n-2)!}
A few comments in this simple case will prove useful later.

\noindent{}1) The fact that the structure constant on the r.h.s.
of \nonlin\ is $1$ was originally obtained by calculating a four
point
correlation function. On general grounds one expects a relation like:
\eqn\gnonl{
a_-\CT_{k_1}^{(+)}\int\CT_{k_2}^{(+)}=
f(k_1,k_2)\CT_{k_1+k_2-2}^{(+)}}
with some apriori unknown structure function $f(k_1,k_2)$. The point
is that one can determine $f(k_1,k_2)$ by requiring associativity of
\gnonl. Indeed, consider:
\eqn\consist{(a_-)^2\CT_{k_1}^{(+)}
\CT_{k_2}^{(+)}
\CT_{k_3}^{(+)}=f(k_1,k_2)f(k_1+k_2-2,k_3)\CT_k^{(+)}=
(k_1\leftrightarrow k_2)=
(k_1\leftrightarrow k_3)}
where $k=k_1+k_2+k_3-4$.
Associativity \consist\ implies that
$f(k_1,k_2)f(k_1+k_2-2,k_3)$
must be invariant under permutations of $k_1,k_2,k_3$.
By applying $a_+$ to \gnonl\ one also finds that
$f(k_1,k_2)$ should be periodic in $k_1, k_2$ (with period $2$). The
solution can be expressed in terms of an arbitrary (periodic)
function $M(k)$: $f(k_1,k_2)=M(k_1+k_2)/M(k_1)M(k_2)$; $M(k)$ can be put to $1$
by renormalizing $\CT_k$. Below we will see less trivial examples
of such structure constants which will also be determined by
associativity
conditions.

\noindent{}2) The relations \mod, \nonlin\ provide a link to the
formalism
of \polc. Consider first \mod; it implies $a_+a_-\CT_k=0$, a fact
that
has been interpreted in \KMS\ as the statement that the $\CT_k$
describe
infinitesimal excitations living on the line $a_+a_-=0$ in the $(a_+,
a_-)$
plane. This line is related to the Fermi surface of the matrix model
fermions through the relations \wkms\ $a_\pm=p\pm q$ ($p,q$ are phase
space variables).
$\CT^{(\pm)}$ live on the segments $a_\mp=0$. Eq. \nonlin\ is very
natural
in this interpretation; indeed, it can be rewritten as:
\eqn\pert{a_-\CT_{k_1}^{(+)}|_{\lambda_{k_2}}=\lambda_{k_2}^{(+)}
\CT_{k_1+k_2-2}^{(+)}}
where $\lambda_k$ is the condensate of $\CT_k$, corresponding to
${\cal S}\rightarrow{\cal S}+\lambda_k^{(\pm)}\int \CT_k^{(\pm)}$ in
\action.
Combining \mod\ and \pert\ we see that perturbing by $\lambda_k$
shifts
the Fermi surface to
\eqn\Ferm{a_+a_-=\lambda_k e^{ikX}}
where as usual \Ferm\ is understood as a relation on the tachyon
module \KMS\ (rather than a relation in the ring).
In particular, for $k=0$ we find \Ferm\ $a_+a_-=\mu$ \wkms.
Equation \Ferm\ is very reminiscent of the results of \polc.

The above discussion is a step towards establishing the connection
with
the matrix model. However the picture is still incomplete; one
difficulty
is that \mod, \pert\ (and obvious generalizations)
do not seem to hold in non bulk amplitudes.
Nevertheless, we will next turn to obtaining the analogs
of \mod\ -- \Ferm\ for the disk with open and closed states.

\newsec{The disk with open external states.}

\subsec{Ring and modules.}

Specializing the spectrum \clop\ to $D=1+1$ we find, in addition to
the
closed tachyon \closed, another massless field, the open string
tachyon:
\eqn\open{\CV_k^{(\pm)}=-\Gamma(\pm k)
  \exp\left[i{k\over2\sqrt2}X+(-{1\over\sqrt{2}}\pm{k\over2\sqrt2})
\phi\right]}
(we have again chosen a convenient normalization). The open string
excitations
form modules of the open string (or boundary) ring, which is
generated by
the operators\foot{The boundary conditions
on the ghosts are standard, $c=\bar c$, etc. We will use
$c,b$ for the ghosts on the boundary; it should be evident from the
context
which of the two $c,b$'s is relevant in different expressions.}:

\eqn\aopen{A_{\pm}=-(cb -
                   {1 \over 2\sqrt 2}(\partial \phi {\mp} i\partial
X))
         e^{{\pm}iX/2\sqrt2 + \phi/2\sqrt2 }  }
One can readily verify that:
\eqn\omod{
\eqalign{
A_\pm \CV_k^{(\pm)}=&\CV_{k\pm1}^{(\pm)}\cr
A_\mp \CV_k^{(\pm)}=&0\cr}}
A new feature of the open case is the importance of the ordering of
the
operators on the boundary of the disk. The
relations \omod\ with opposite ordering are easily obtained
by using the symmetry $\sigma\rightarrow-\sigma$, under which
\aopen\ $A_\pm\rightarrow-A_\pm$ while the tachyons are invariant.
Hence
for example, $\CV_k^{(\pm)}A_\pm=-\CV_{k\pm1}^{(\pm)}$. We will use
this symmetry below and write only the independent relations in each
case.

\subsec{Correlation functions.}

In the next section we will consider scattering amplitudes with an
arbitrary
number of closed \closed\ and open \open\ states. However, as a
useful
intermediate step, we will start with a discussion of open string
scattering, since this will allow us to exhibit the new features
arising
in generalizing \FZ\ -- \Ferm\ to the disk.

A natural starting point is the amplitude:
\eqn\FS{F(\sigma)=\langle
\prod_{j=1}^m\CV_{p_j}^{(-)} A_-(\sigma) \prod
_{i=1}
^n\CV_{k_i}^{(+)} \rangle}
By \FS\ we mean the {\it ordered} amplitude on the disk with the
specified
ordering.
To make sense out of \FS\ one has to fix three of the $\CV^{(\pm)}$
in \FS\ and integrate over the rest. It would seem, from the way
we wrote \FS, that $F(\sigma)$ is independent of which three
operators
we fix. This is {\it not} the case. Different ways of fixing\foot{More
precisely, the $SL(2,R)$ invariant correlation function is
$\int\partial_\sigma F(\sigma)$.}
$SL(2,R)$
give rise to different $F(\sigma)$, but the constraints on tachyon
amplitudes
\venez\ that we will derive are of course independent of this choice
(as are the original amplitudes \venez).
We will start with a particularly convenient way of
fixing
$SL(2,R)$ and briefly describe other ways below (and present more
details
in an appendix).

One way of defining \FS\ is:
\eqn\FSC{F(\sigma)=\langle \prod_{j=1}^{m-1}\int\CV_{p_j}^{(-)}
c\CV_{p_m}^{(-)}(0)A_-(\sigma)c\CV_{k_1}^{(+)}(1) \prod_{i=2}
^{n-1}\int\CV_{k_i}^{(+)} c\CV_{k_n}^{(+)}(\infty)\rangle}
The reason why \FSC\ is convenient is that $\partial_\sigma
F(\sigma)=0$
in this choice. Naively, this should always be the case since
\eqn\opBRS{\partial_\sigma A_-(\sigma)=\{Q_\brs,b_{-1} A_-(\sigma)\}}
But in general there are finite boundary terms in the moduli integrals
in \FS\
and $\partial_\sigma F(\sigma)\not=0$ (see below and appendix A). In
\FSC\
such boundary terms vanish (as one can verify
by explicit calculation), therefore we can argue as in \FZ\ --
\sprec; compare $F(\sigma\rightarrow0)$ to $F(\sigma\rightarrow1)$.
In the
first case we find \omod: $-\langle\prod_{j=1}^{m-1}\CV_{p_j}^{(-)}
\CV_{p_m-1}^{(-)}\prod_{i=1}^n\CV_{k_i}^{(+)}\rangle$.
To act to the right $(\sigma\rightarrow1)$ we have to understand the
equivalent of \nonlin\ here. This is readily done; one finds:
\eqn\first{A_-\CV_{k_1}^{(+)}\int\CV_{k_2}^{(+)}={1\over\sin\pi k_1}
\CV_{k_1+k_2-1}^{(+)}}
Of course, by $SL(2,R)$ invariance, the same holds for
$A_- \int\CV_{k_1}^{(+)}
\CV_{k_2}^{(+)}$. Since ordering is important,
there is a second independent relation which is not necessary here but
will be useful below:
\eqn\second{\int\CV_{k_1}^{(+)}A_-\CV_{k_2}^{(+)}={\sin\pi(k_1+k_2)
\over\sin\pi k_1\sin\pi k_2}
\CV_{k_1+k_2-1}^{(+)}}
As explained above, there are two more implicit relations in \first,
\second:
$$\CV_{k_1}^{(+)} \int\CV_{k_2}^{(+)} A_-=-{1\over\sin\pi k_2}
\CV_{k_1+k_2-1}^{(+)};\;\;
\CV_{k_1}^{(+)} A_- \int\CV_{k_2}^{(+)}=-
{\sin\pi(k_1+k_2)
\over\sin\pi k_1\sin\pi k_2}
\CV_{k_1+k_2-1}^{(+)}~.$$
We are now ready to return to \FSC. Using \first\ in
$F(\sigma\rightarrow1)$
and comparing to
$F(\sigma\rightarrow0)$
we find the following recursion relation:
\eqn\recur{\langle \prod_{j=1}^{m-1}\CV_{p_j}^{(-)}
\CV_{p_m-1}^{(-)} \prod_{i=1}^n\CV_{k_i}^{(+)}
\rangle=-{1\over\sin\pi
k_1}
\langle \prod_{j=1}^{m}\CV_{p_j}^{(-)}
\CV_{k_1+k_2-1}^{(+)} \prod_{i=3}^n\CV_{k_i}^{(+)} \rangle}
which is easily solved:
\eqn\ans{\langle \prod_{j=1}^m\CV_{p_j}^{(-)}
\prod_{i=1}^n\CV_{k_i}^{(+)}\rangle
=(-1)^{{n(n-1)\over2}+{m(m-1)\over2}}
\prod_{r=1} ^{m-1} {1 \over \sin\pi \sum_{j=1}^r p_j}
\prod_{s=1}^{n-1} {1 \over \sin\pi \sum_{i=1} ^s k_i}~,}
in agreement with the answer obtained in \bk.
Similarly one can check that as claimed in \bk\ the amplitudes for more
complicated orderings of $\CV_+,\CV_-$ vanish.
We see that as on the sphere,
the correlation functions are determined by the ring action on the tachyon
modules
\omod, \first.

\subsec{Other ways of fixing SL(2,R).}

The reader may have noticed a peculiar feature of the derivation
of \recur. Unlike the closed case, since ordering is important here,
the result
\recur\ seems to depend strongly on the fact that one of the vertices
$\CV_{k_1}^{(+)}$,
$\CV_{k_2}^{(+)}$ in \FS\ is fixed, while the other one is integrated
(allowing us to use one of the two versions of \first). Naively
one would get a different answer if they were both fixed or both
integrated.
The resolution of this `paradox' is that in those cases, as in other
important cases below, $\partial_\sigma F(\sigma)\not=0$. As noted
above,
\opBRS, $\partial_\sigma F(\sigma)$ is a correlation function with an
insertion
of a BRST commutator, and naively it should vanish. However, one can show
that
precisely in the cases just mentioned there are finite contributions
from the boundaries of moduli space which precisely complete \recur.
The reader
should consult appendix A for a more detailed discussion of this
phenomenon.

We would like to emphasize, that the fact that the BRST commutator
\opBRS\ does not decouple in general $(\partial_\sigma F(\sigma)\not=0)$ does
not
imply breakdown of gauge invariance of the theory. The amplitude \FS,
\FSC\
is not the full scattering amplitude -- one has to sum over different
orderings of the vertices. After doing the sum, one can show that BRST
invariance is restored. For our purposes it is more convenient to
consider the individual `channels'.

\subsec{The algebraic structure.}

The action of the ground ring on the tachyon modules is, as in the
closed
case, highly constrained. Indeed, replace \first, \second\ by:
\eqn\general{
\eqalign{
A_-\CV_{k_1}^{(+)}\int\CV_{k_2}^{(+)}=&f(k_1,k_2)
\CV_{k_1+k_2-1}^{(+)}\cr
 \int\CV_{k_1}^{(+)} A_-\CV_{k_2}^{(+)}=&g(k_1,k_2)
\CV_{k_1+k_2-1}^{(+)}\cr}}
Consider the operator
$\Theta= \int\CV_{k_1}^{(+)}
A_-\CV_{k_2}^{(+)} \int\CV_{k_3}^{(+)} A_-$.
We can calculate $\Theta$ in two different ways by using \general.
Comparing the coefficients of $\CV_{k_1+k_2+k_3-2}$ we obtain a consistency
relation on $f, g$:
\eqn\jac{\eqalign{f(k_3,k_2)g(k_1,k_2+k_3-1)&=\cr
           f(k_3, k_1+k_2-1) g(k_1, k_2) +&
                    f(k_2,k_3) f(k_2+k_3-1, k_1) \cr}}
The functions $f,g$ are periodic in $k_1, k_2$ with period $1$
(for the same reasons as in \gnonl, \consist). One can also show that the
condition \jac\ together with periodicity determine $f,g$ essentially
uniquely. It is
a non trivial fact that
$f(k_1, k_2)={1\over\sin\pi k_1}$ and
$g(k_1,k_2)={\sin\pi(k_1+k_2)\over\sin\pi k_1\sin\pi k_2}$
indeed satisfy \jac\
as well as an infinite number of more complicated
consistency conditions implied by \general. This gives rise to some
quite non trivial trigonometric identities.

Another way of probing the consistency of the structure we found \omod,
\first, \second\ is to consider the action of the ring in more complicated
correlation functions (than \FS). Consider for example:
\eqn\GS{G(\sigma)=
\langle\cdots\CV_{k_1}^{(+)}(0) \int\CV_{k_2}^{(+)} A_-(\sigma)
\CV_{k_3}^{(+)}(1) \int\CV_{k_4}^{(+)} \cdots\rangle}
where the $\cdots$ stand for other operators which may be present.
One can show that $\partial_\sigma G(\sigma)=0$, and therefore,
$G(\sigma\rightarrow0)=
G(\sigma\rightarrow1)$.
Using the ring relations
\general\ one gets an equation relating three amplitudes
which can be viewed as a consistency condition on $f,g$. That condition
is equivalent to \jac.
Similar manipulations with higher powers of $A_-$ give rise to
the infinite number of associativity conditions mentioned above.

\newsec{Scattering of Open and Closed Strings.}

\subsec{Algebraic structure and Recursion relations.}

We now turn to the scattering amplitudes with an arbitrary number of open
and closed states. Following the logic of section 3 we will obtain recursion
relations by studying the ring relations for the amplitude:
\eqn\Fopcl{C(\sigma)=\langle \prod_{j=1}^m\CV_{p_j}^{(-)} A_-(\sigma)
 \prod_{i=1}
^n\CV_{k_i}^{(+)}   \prod_{a=1}^M \CT_{r_a} ^{(-)}
 \prod_{s=1}^N
\CT_{q_s} ^{(+)} \rangle}
In the previous section (and appendix A) we have discussed the case
$N=M=0$, and saw that the ring relations on the (open) tachyon modules
\omod, \first, \second\ determine
the correlation functions \Fopcl\ uniquely. In the general case $(N,M\not=0)$
the same conclusion holds. The only new element is the action
of $A_\pm$ on the closed tachyons $\CT^{(\pm)}$. The relevant
calculations are presented in appendix B, where we show that $\partial_\sigma
C(\sigma)\not=0$ (for any choice of $SL(2,R)$ fixing);
for the particular choice used in section 3 \FSC\ the relevant boundary
terms are due to closed tachyons ($\CT_{q_i}^{(+)}$) approaching
$A_-$. They can be summarized by the following action of the boundary ground
ring on the closed tachyons:
\eqn\Bcl{
\eqalign{
A_- \CT_{q}^{(+)}=& \sin ({\pi q \over2})
\CV_{q-1}^{(+)}\cr
A_- \CT_{q}^{(-)}=&0\cr}}
The structure constant in \Bcl\ is again determined by consistency
with the other relations \omod, \first, \second\ as in section {\it 3.4}.
The details of how \Bcl\ emerges from \Fopcl\ appear in appendix B, but
combining it with the previous relations one can immediately write
a recursion relation for the general amplitude \Fopcl:
\eqn\recOC{\eqalign{&\langle \prod_{j=1}^{m-1}\CV_{p_j}^{(-)}
\CV_{p_m-1}^{(-)} \prod_{i=1}^n\CV_{k_i}^{(+)}
 \prod_{a=1}^M \CT_{r_a}
^{(-)}    \prod_{s=1}^N \CT_{q_s} ^{(+)}  \rangle=\cr
&~-{1\over\sin\pi k_1}  \langle \prod_{j=1}^{m}\CV_{p_j}^{(-)}
\CV_{k_1+k_2-1}^{(+)} \prod_{i=3}^n\CV_{k_i}^{(+)}
 \prod_{a=1}^M
\CT_{r_a} ^{(-)}    \prod_{s=1}^N \CT_{q_s} ^{(+)} \rangle \cr
&~-\sum_{i=1} ^N \sin {\pi q_i \over2}
\langle \prod_{j=1}^{m}\CV_{p_j}^{(-)}
\CV_{q_i-1}^{(+)} \prod_{i=1}^n\CV_{k_i}^{(+)}
 \prod_{a=1}^M \CT_{r_a}
^{(-)}    \prod_{i\neq s=1 } ^N \CT_{q_s} ^{(+)} \rangle
}}
The first term on the r.h.s. of \recOC\ vanishes for $n<2$, while the second
vanishes for $N=0$. Eq. \recOC\ generalizes \recur\ by expressing
a general correlation function of open and closed states (l.h.s.)
in terms of one with one fewer open string state (the first term
on the r.h.s.) and another with one fewer closed string state (second
term on r.h.s.).

A few comments about \recOC\ are in order:

\noindent1) It clearly determines all correlation functions uniquely. By
iterating it enough times, an arbitrary correlation function \Fopcl\
is related to the (known) three point function
$\langle \CV\CV\CV\rangle$.

\noindent2) The dynamics splits into that of left and right
moving particles in space time; the solution to \recOC\ has the form:

\eqn\decoup{\langle \prod_{j=1}^{m}\CV_{p_j}^{(-)}
 \prod_{i=1}^n\CV_{k_i}^{(+)}
 \prod_{a=1}^M \CT_{r_a}
^{(-)}    \prod_{s=1}^N \CT_{q_s} ^{(+)}  \rangle=
W_{n,N}  (k_i| q_s) W_{m,M} (p_j|r_a)}
where $W_{n,N}$ is a certain function of momenta determined by the recursion
relation \recOC\ and
the only coupling between left and right being through the zero mode
sum rules:
$$\sum_{i=1}^nk_i+\sum_{s=1}^Nq_s=-
\sum_{j=1}^mp_j-\sum_{a=1}^Mr_a=n+m+2(N+M)-2$$
This generalizes a similar structure observed in \bk\ for the case $N=M=0$
in \decoup.

\noindent3) The solution to \recOC\ clearly has the periodicity
$k_i\rightarrow k_i+1; q_s\rightarrow q_s+2$
(and similarly for the left movers), again generalizing \bk.

\subsec{The form of the solutions}

One can solve \recOC\ more explicitly in terms of known functions.
The dynamics of the right (left) moving particles is governed by
 $W_{n,N}(k_i|q_s)$ ($W_{m,M}(p_i|r_a)$). In order to compute
$W_{n,N}(k_i|q_s)$
consider the
partition of all  $N$ left moving  closed tachyons into $n$ sets ($n$
is the number of left moving open tachyons):
\eqn\part{{\cal P}=\lbrace q_1,q_2,...q_N \rbrace=\oplus_{i=1} ^{n}
\lbrace q_{j_1}^{(i)}...q_{j_{v_i}}^{(i)}\rbrace~, }
where $v_i$ is the number of momenta in the $i$-th set.
Some of the $v_i$ may be equal to zero, in which case the
set is empty. Then
$W_{n,N}^{(+)}(k_i,q_s)$ is given by the sum over all partitions $\cal P$
\part\ in terms of the basic functions $W_{n,0}$, $W_{1,r}$:
\eqn\anzatz{W_{n,N}(k_i|q_s)=\sum_{\cal P}\left(
         \prod_{i=1} ^{n} W_{1,v_i}
(-\sum_{s=1} ^{v_i} q_{j_s}^{(i)}|q_{j_s}^{(i)})\right)
     W_{n,0} (k_i+\sum_{s=1} ^{v_i} q_{j_s}^{(i)}|0) }
$W_{n,0}$ is known from the open case (compare \ans, \decoup), while
$W_{1,r}$
satisfies a simple recursion relation:
\eqn\simple{W_{1,r}^{(+)}(k|q_i)=
\sum_{j=1}^r {\sin \pi (k +{q_j \over 2}) \over \sin \pi k }
W_{1,r-1}^{(+)}(k+q_j|q_i,\not q_j)~,}
where $k$ and $q_i$ are related by momentum conservation.
The relation  \simple\
should be supplemented with the initial condition $W_{1,0}=1$.
Its derivation follows from the ring relations
(using $A_-^2$ instead of $A_-$ in \FS); we will omit the details
of the proof.

The simplest example of \anzatz\ is the scattering amplitude of one
closed string
tachyon and an arbitrary number of open tachyons. Taking into account that
$W_{1,1}(-q|q)=(2\cos{\pi q \over 2})^{-1}$ we get
\eqn\oneCl{\eqalign{\langle & \prod_{j=1}^{m} \CV_{p_j}
^{(-)}   \prod_{i=1}^n\CV_{k_i}^{(+)}
\CT_q ^{(+)} \rangle = \cr
   {1 \over 2\cos {\pi q \over 2 }} \sum_{l=1}^n &\langle
       \prod_{j=1}^{m} \CV_{p_j} ^{(-)}    \prod_{i=1}^{l-1}
\CV_{k_i}^{(+)}
 \CV_{k_l+q}^{(+)}   \prod_{i=l+1} ^n \CV_{k_i}^{(+)}   \rangle \cr}}

\newsec{Summary.}

In this paper we have presented the ring -- module structure of open plus
closed string theory on the disk. We have seen that the action of the
ground ring on the modules both linear \mod, \omod, and non-linear
\nonlin, \first, \second, \Bcl\ is completely determined by
consistency of the structure and furthermore determines the correlation
functions uniquely through the basic relation \recOC\ which follows
from the ring structure.
The ring relations are powerful and we have been able to solve for
the correlation functions without doing any integrals. This led to a
verification of known results for the correlation functions of open string
tachyons \ans\ and to new results for correlation functions of open and
closed tachyons \recOC, \anzatz. The algebraic
structure of the disk correlation functions is quite rich and we have certainly
not exhausted it here. For higher genus Riemann surfaces one should be able
to make progress in a similar fashion
by generalizing the ring relations. It is important to study this structure
(and possible generalizations)
in more detail.

The action of the ring on the modules is suggestive again of a phase space
interpretation with excitations living on certain curves. It is interesting
that both open and closed tachyons seem to live on the same $(A_+, A_-)$
phase space. Naively $(a_+, a_-)$ generate independent directions,
but this is probably not the case -- there should be relations of the general
form $a_\pm\simeq A_\pm^2$ (in the sense of \KMS\ -- acting on the
modules).
One would also like to extend the results of this paper to non-bulk
amplitudes and try to solve the relevant matrix model
(see e.g. \ref\kazkos{V. Kazakov and I. Kostov, Paris preprint (1992).})
to gain further
insight on the physics of these models, which is bound to be interesting.

\bigbreak\bigskip\bigskip\centerline{{\bf Acknowledgements}}\nobreak

We thank C. Vafa and A. B. Zamolodchikov for discussions.
This work was partially supported by
DOE grant DE-AC02-76ER-03072 and
by Packard Fellowship 89/1624.

\appendix{A}{Open tachyon scattering.}

In the text we have extracted a recursion relation \recur\ for the
scattering amplitude of open string tachyons by considering
a correlation function involving an $A_-$ \FSC.
We have stressed that the derivation of \recur\ depended crucially
on the way we defined \FS\ in \FSC. The purpose of this appendix
is to show that different ways of defining \FS\ give rise to
the same recursion relation \recur, but in a
different way.
Consider:
\eqn\FSCA{\tilde F(\sigma)=\langle \prod_{j=1}^{m-1}\int\CV_{p_j}^{(-)}
c\CV_{p_m}^{(-)}(0)A_-(\sigma)
c\CV_{k_1}^{(+)}(1)
c\CV_{k_2}^{(+)}(\infty)
 \prod_{i=3}^n
\int\CV_{k_i}^{(+)} \rangle}
$\tilde F$ is {\it not} equal to $F(\sigma)$ \FSC.
As in the text, we are interested in computing $\int_0^1\partial_\sigma
\tilde F(\sigma)$. One way to calculate it is to use the
fact that it is clearly equal to $\tilde F(\sigma=1)-\tilde F(\sigma=0)$.
Comparing to \FSC\ we see that the two $\CV^{(+)}$'s closest to $A_-$
are not integrated, hence there is naively no contribution
of the form \first\ from $\sigma=1$. This is indeed correct (but still
naive as we'll see in a moment), and therefore,
\eqn\DF{\int_0^1\partial_\sigma \tilde F(\sigma)=-\tilde F(\sigma=0)=
-\langle\prod_{j=1}^{m-1}\CV_{p_j}^{(-)}
\CV_{p_m-1}^{(-)}\prod_{i=1}^n\CV_{k_i}^{(+)}\rangle}
On the other hand, we may compute $\partial_\sigma \tilde F(\sigma)$ by using
$\partial_\sigma A_-(\sigma)=\{Q_\brs, b_{-1}A_-\}$ \opBRS;
commuting $Q_\brs$ through the rest of the operators in \FSCA\ we pick
up potential boundary contributions from each of the integrated $\CV_i$:
$\int\partial(c\CV_i)$. The main question is whether there are finite
boundary terms. One can convince oneself that the situation is the following:
for all the $\CV_{p_j}^{(-)}$ $j=1,..., m-1$ and $\CV_{k_i}^{(+)}$
$i=4,...,n$ there are finite and equal boundary contributions
from the upper and lower limits of integration, which therefore cancel
each other. For $\CV_{k_3}^{(+)}(z_3)$ the analysis is slightly modified:
the upper limit of integration $(z_3\rightarrow\infty)$ gives a vanishing
boundary term, whereas the contribution from $z_3\rightarrow0$ is finite.
The value of the boundary term from $z_3\rightarrow0$ is given by
a product of correlation functions:
$$\langle\int b_{-1}A_-\CV_{k_1}^{(+)}\CV_{k_2}^{(+)}\CV_k^{(+)}\rangle
\langle\prod_{j=1}^m\CV_{p_j}^{(-)}\CV_{-k}^{(+)}\prod_{i=3}^n
\CV_{k_i}^{(+)}\rangle$$
which precisely supplies the missing term to turn \DF\ into \recur.
The relation between the way \recur\ arises from
\FSC\ and \FSCA\ is through a conformal transformation;
in \FSC\ we allow $A_-$, $\CV_{k_1}^{(+)}$ and $\CV_{k_2}^{(+)}$
to approach each other, and get a finite contribution
from the region of degeneration;
in \FSCA\ we keep the distance between
$\CV_{k_1}^{(+)}$ and $\CV_{k_2}^{(+)}$
fixed and the degeneration region is the region  where all other
operators approach each other (while being well separated from $\CV_{k_1}
^{(+)}, \CV_{k_2}^{(+)}$).
We see that the result of this calculation can be summarized by the additional
relation:
$$A_-\CT_{k_1}^{(+)}\CT_{k_2}^{(+)}={1\over\sin\pi k_1}\CT_{k_1+k_2-1}^{(+)}$$
and appropriately for \second.

\appendix{B}{Open and closed tachyon scattering.}

Here we supply some details of the derivation of eq. \Bcl, \recOC\
which determine the open plus closed string correlation functions.
Consider the function:
\eqn\CS{C(\sigma)=\langle\prod_{j=1}^{m-1}\int\CV_{p_j}^{(-)}
c\CV_{p_m}^{(-)}(0)A_-(\sigma)c\CV_{k_1}^{(+)}(1)\prod_{i=2}
^{n-1}\int\CV_{k_i}^{(+)}c\CV_{k_n}^{(+)}(\infty)
\prod_{a=1}^M \int\CT_{r_a} ^{(-)}
\prod_{s=1}^N \int
\CT_{q_s} ^{(+)}\rangle~,}
where the $\CV$ are integrated over the real line in the specified order,
while the $\CT$ are integrated over the whole upper half plane.
In the by now standard fashion we consider $\int_0^1\partial_\sigma
C(\sigma)$ and compute it in two different ways. The first, as $C(\sigma
\rightarrow1)-C(\sigma\rightarrow0)$ gives rise using \omod, \first\
to the standard terms in \recOC: the l.h.s. and the first term on the r.h.s..
The second, obtained again by writing $\partial_\sigma A_-$ as a BRST
commutator
\opBRS\ reduces to a sum of total derivative terms from each of the integrated
vertices in \CS.
Many of these terms vanish. In particular, there are no boundary contributions
from $\CV_{p_j}^{(-)}$,
$\CV_{k_i}^{(+)}$,
$\CT_{r_a}^{(-)}$. Each of the $\CT_{q_s}^{(+)}$ gives rise
to a boundary contribution from the region near $\sigma$:
$\int\partial_z(c\CT_{q_s}^{(+)})b_{-1}A_-(\sigma)$, which can be
written as:
\eqn\Form{\langle\CT_{q_s}^{(+)}b_{-1}A_-\CV_k^{(+)}\rangle\langle\CV_{-k}
^{(+)}\cdots\rangle}
This is the origin of the second term on the r.h.s. of \recOC.
To actually calculate the structure function $(\sin{\pi q_s\over2})$
in \Bcl, \recOC\ one can either explicitly evaluate the correlator
\Form, or apply various consistency requirements, as mentioned above.

\listrefs
\end